\documentclass[onecolumn]{gji}
\usepackage{timet,color}
\usepackage[urlcolor=blue,citecolor=black,linkcolor=black]{hyperref}
\usepackage{float}
\usepackage{graphicx}
\graphicspath{ {./figures/} }
\usepackage{lipsum}
\usepackage{mathtools}
\usepackage{cuted}

\title[Attenuation of surface modes in granular media]
  {Geometric and material attenuation of surface acoustic modes in granular media}
\author[R. Zaccherini et al.]
  {R. Zaccherini$^1$, A. Palermo$^2$, A. Marzani$^2$, A. Colombi$^1$, V. K. Dertimanis$^1$, E. N. Chatzi$^1$ \\
  $^1$ Department of Civil, Environmental and Geomatic Engineering, ETH Z\"urich, Z\"urich 8093, Switzerland. \\ E-mail: zaccherini@ibk.baug.ethz.ch \\
  $^2$ Department of Civil, Chemical, Environmental and Materials Engineering - DICAM, University of Bologna, \\ Bologna 40136, Italy
  }
  
\pagerange{\pageref{firstpage}--\pageref{lastpage}}
\volume{--}
\pubyear{2021}

\begin{document}

\label{firstpage}

\maketitle

\begin{summary}
 In this work, an unconsolidated granular medium, made of silica microbeads, is experimentally tested in a laboratory setting. The objective is to investigate the attenuation mechanisms of vertically polarized seismic waves traveling at the surface of unconsolidated substrates that are characterized by power-law rigidity profiles. Both geometric spreading and material damping due to skeletal dissipation are considered. An electromagnetic shaker is employed to excite the granular medium between 300 and 550 Hz, generating linear modes that are localized near the surface. A densely sampled section is recorded at the surface using a laser vibrometer. The explicit solution of the geometric attenuation law of Rayleigh-like waves in layered media is employed to calculate the geometric spreading function of the vertically polarized surface modes within the granular material. In accordance with recent studies, the dynamics of these small-amplitude multi-modal linear waves can be analysed by considering the granular medium as perfectly continuous and elastic. By performing a non-linear regression analysis on particle displacements, extracted from experimental velocity data, we determine the frequency-dependent attenuation coefficients, which account for the material damping.\\
 The findings of this work show that laboratory-scale physical models can be used to study the geometric spreading of vertically polarized seismic waves induced by the soil inhomogeneity and characterize the material damping of the medium.

\end{summary}


\section{Introduction}
Granular materials, including natural sand, silica spheres, and glass beads, have garnered increasing interest within the geophysics and geotechnical communities, as they enable the physical and geological modeling of various complex structures \cite{Sherlock,Krawczyk}. This is due to the wide range of shapes and sizes of their grains, as well as their intrinsic mechanical parameters. 
In particular, glass microbeads (GBs) have recently been employed in laboratory-scale physical models to study the propagation of seismic waves in unconsolidated \cite{Jacob,Bodet,Bergamo} and porous \cite{Pasquet} heterogeneous media. Small-scale physical modelling, combined with laser measurement techniques, is often used to address theoretical and methodological issues related to seismic exploration, especially when experimental tests or inversion techniques are needed \cite{Campman,Bodet3,Bodet4,Dewangan,Bretaudeau}.\\
Under gravity loads, the elastic properties of granular materials can be defined according to the Hertzian theory, which describes the contact forces between the beads \cite{Gassmann}. The theory predicts that, near the free surface, the longitudinal $v_P$ and shear $v_S$ wave velocity profiles exhibit a power-law dependence on the compacting pressure $p=\rho gz$ along the depth $z$, where $\rho$ is the medium density and $g$ the gravitational constant. Consequently, the wave velocity profiles of a granular material can be expressed as $v_{P,S} = \gamma_{P,S} (\rho g z)^{\alpha_{P,S}}$, with $\gamma_{P,S}$ designating a coefficient that is mainly influenced by the medium elastic properties and porosity and $\alpha_{P,S}$ denoting the power-law exponent; estimated to equal 1/6, when uniform beads randomly arranged in a close-packed structure are considered. Whether this coefficient is correct when describing actual granular materials is still debated \cite{Makse}. Recent experiments suggest that imperfections in the geometry or dimension of the beads might affect the value of $\alpha_{P,S}$. Regardless of the physical origins of such a variation in the power-law coefficients, it is worth noting the consequences this has on the acoustic wave propagation. In particular, the increase in material stiffness with depth causes the upward bending of the acoustic waves (also known as the ``mirage" effect \cite{Liu}) toward the free surface \cite{Gusev2}. Thus, the depth-increasing stiffness profile of the medium combined with the presence of a mechanically free surface leads to the formation of guided surface acoustic modes (GSAMs) channeled between the free surface and the increasingly rigid material \cite{Gusev,Bonneau}. Among these GSAMs are vertically polarized P-SV waves, composed of coupled longitudinal (P) and shear vertical (SV) modes \cite{Aleshin}. Experimental dispersion analysis of these P-SV acoustic modes propagating along the surface of granular media has been conducted in existing literature, with the purpose of estimating the elastic power-law coefficients \cite{Jacob,Bodet2,Bodet}. In addition, experimentally determined dispersion relations have revealed unusual low propagation velocities, ascribed to very low pressures of the near-surface layers \cite{Jacob}.\\
While the dynamics of the P-SV waves has been extensively investigated, the attenuation of their amplitude due to geometric spreading and material damping remains uncharted. The geometric spreading is mainly controlled by the source type (e.g., point or line source) and the elastic profile of the medium \cite{Foti}, whereas the material (or internal) damping is linked to the energy dissipated within the skeletal frame of the medium \cite{Rix}. Both geometric spreading and material damping strongly influence the seismic site response. Determination of attenuation characteristics of shallow unconsolidated soil layers is fundamental to the study of several problems in geophysics and geotechnical engineering, such as site amplification or ground borne vibrations \cite{Field}. The objective of this work is to determine the P-SV wave geometric spreading in order to gain insights into the attenuation mechanisms of vertically polarized waves traveling at the surface of unconsolidated substrates with power-law rigidity profiles at the seismic scale. In addition, frequency-dependent P-SV wave attenuation coefficients are determined to characterize the material damping of the employed granular medium.\\
In geophysics, surface wave methods (SWMs) are among the most established in situ test methods for the evaluation of geometric and material attenuation properties of near-surface soils \cite{Rix,Foti}. The SWMs include the spectral slope \cite{Jongmans,Stewart,Gibbs}, waveform inversion \cite{Askan}, half-power bandwidth \cite{Badsar}, circle fit \cite{Verachtert} techniques, and the regression of the amplitude of particle motion as a function of the distance from the source, among others.\\
In this work, the particle motion amplitude regression technique is used to determine the attenuation properties of a granular layer of GBs. The amplitude decay of the displacement associated with the P-SV surface modes is analysed as the wavefront propagates away from the source. We replicate the experimental setup designed in Refs. \cite{Jacob,Bodet,Bergamo}, which consists of a wooden box filled with an unconsolidated granular material made of silica microbeads. A mechanical point source and a laser-Doppler vibrometer are employed to record a small-scale seismic line at the surface. After recalling the dispersion properties of vertically polarized surface modes detected in the experiment, we analyse the effects of geometric spreading and material damping on the multimode P-SV wave displacement. According to the GSAM theory described in Refs. \cite{Aleshin,Jacob}, the P-SV acoustic modes are modeled as linear modes and the granular medium is treated as continuous and elastic, featuring a depth-increasing rigidity gradient. In virtue of the linearity of these modes, the geometric spreading is calculated by adopting an approach originally developed to quantify the spatial attenuation law of seismic Rayleigh-like waves in inhomogeneous stratified media \cite{Foti}. Next, by performing a regression analysis on the particle displacements, as calculated from the experimental velocity data, we determine the frequency-dependent attenuation coefficients, which account for the intrinsic energy dissipation induced by material damping.

\section{Sample preparation and experimental setup}
\begin{figure}
\includegraphics[width=0.5\textwidth]{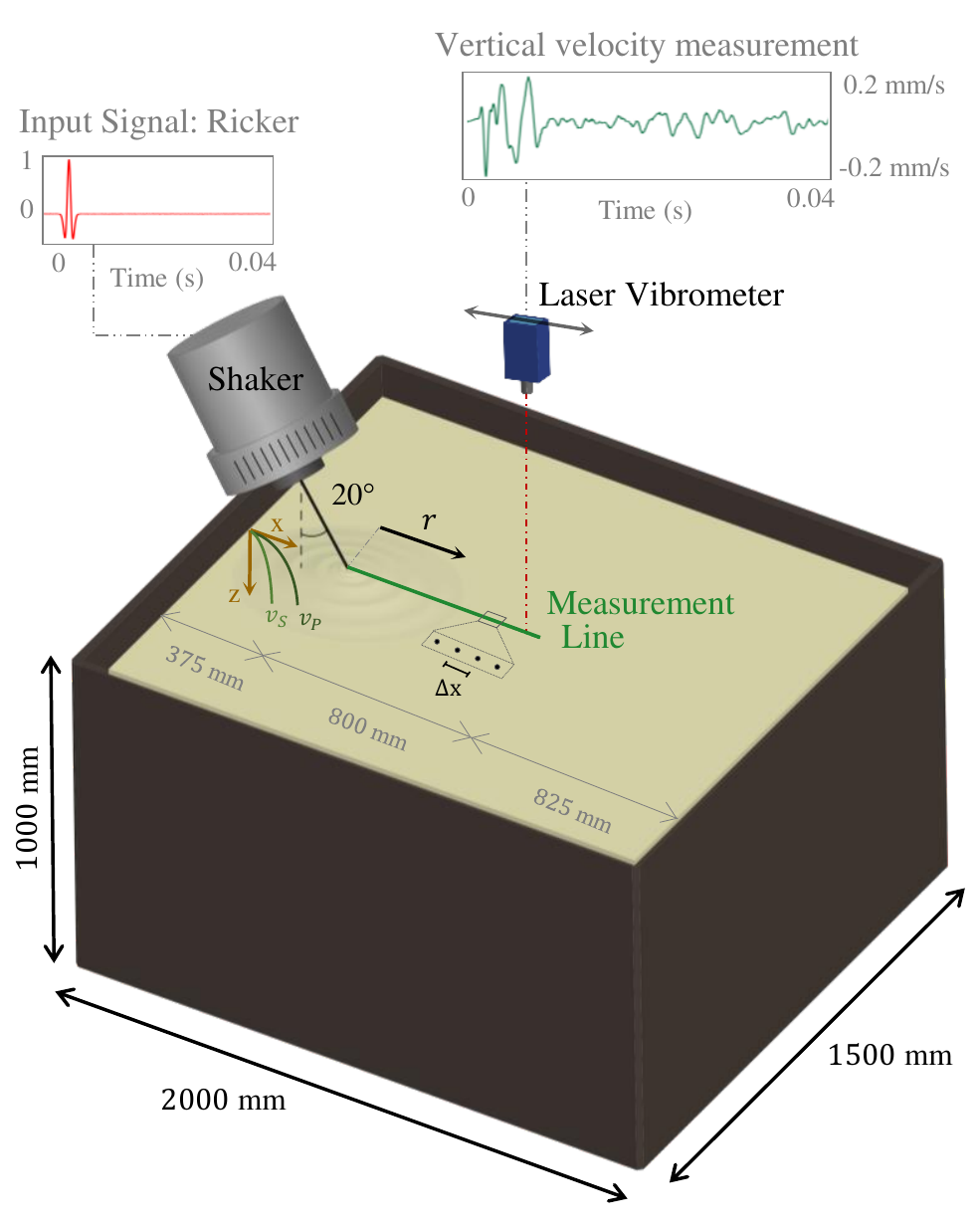}
  \caption{The experimental setup including a wooden box filled with granular material, a laser-Doppler vibrometer, and an electromagnetic shaker driven by a waveform generator, which produces a Ricker pulse centered at 500 Hz. The laser vibrometer, mounted on a scanning stage, records the particle vertical velocity along the 800 mm-long middle line of the box (green line), with constant steps of $\Delta x=6.6$ mm.}
     \label{ExpSetup}
\end{figure}

The experimental setup, illustrated in Fig.~\ref{ExpSetup}, consists of a wooden box of dimensions 2000 ${\times}$ 1500 ${\times}$ 1000 mm filled with an unconsolidated granular material, 150 ${\mu} $m-diameter silica microbeads. To minimize the amplitude of the reflected signals, a paperboard layer is arranged at the bottom of the wooden box. This setup has been originally designed to study the propagation of the GSAMs in granular media \cite{Jacob,Bodet,Bergamo}, while similar ones have been built to analyse their interaction with arrays of mechanical sub-wavelength oscillators \cite{Zaccherini1,Zaccherini2}.\\
The deposition process of the silica particles is designed to ensure a good homogeneity of the medium compaction and, at the same time, estimate its bulk density. Following the procedure adopted in Ref. \cite{Bodet}, the glass beads (approximately 3800 ${kg}$) are gently poured into the wooden box, step by step, in packages of 25 ${kg}$. The thickness of the obtained granular layer is measured to estimate the medium bulk density $\rho$, which is found to be 1600 ${kg/m^{3}}$. 

\noindent
We generate vertically polarized surface waves through a mechanical point excitation at the surface of the medium, situated at a distance of 375 mm from the box edge, and we record the particle vertical velocity field using a Laser-Doppler vibrometer (Polytec, OFV-500). The mechanical point excitation is realized by means of an 8 mm-diameter metal rod in contact with the surface forming an angle of 20$^{\circ}$ with the normal vector to the surface of the granular medium. The rod is attached to an electromagnetic shaker (Tira, Vibration Test Systems N1000 to N2700), driven through a waveform generator (Agilent, 33220), which produces the excitation input signal: a Ricker wavelet centered at 500 Hz. The vibrometer, mounted on a scanning stage moving along a two-dimensional plane parallel to the surface, records the particle vertical velocity along the 800 mm-long middle line of the box, with constant steps of $\Delta x=6.6$ mm (see the green line of Fig.~\ref{ExpSetup}). The chosen step allows for acquisition of 9 points for the shortest wavelength. Following the data acquisition procedure described in Refs. \cite{Jacob,Bodet}, for each point, we average 32 signals of 0.5 s duration at a sampling rate of 500 kHz. Following the procedure detailed in Ref. \cite{Jacob}, all measurements are carried out with small amplitude excitations to ensure that the applied acoustic pulses do not affect the state of the sample. In particular, in the far-field, the wave displacement amplitude is kept below 2 $\mu$m, considerably below the size of the grains (150 $\mu$m). In addition, to limit the effect of non-linearities between the metal rod and the granular material, the maximum vertical velocity close to the source is kept below 3 $\times 10^{-4}$ m/s (in accordance with the procedure described in Ref. \cite{Jacob}, where the maximum vertical velocity was kept below 5 $\times 10^{-4}$). 

\section*{Experimental and numerical dispersion analysis}
\subsection{Experimental data and dispersion analysis}
\begin{figure*}
\includegraphics[width=1\textwidth]{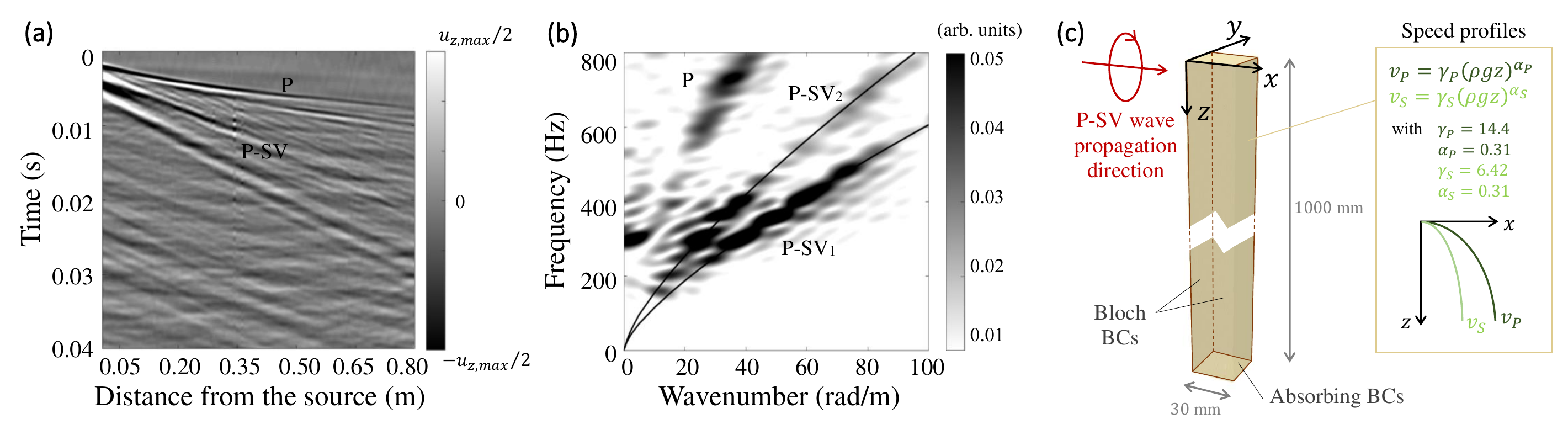}
  \caption{Dispersion Analysis. (a) Seismogram of a Ricker pulse propagating in the granular medium. (b) Experimental and numerical dispersion relations of P-SV acoustic surface modes propagating at the surface of an unconsolidated granular medium. (b) A drawing of the 3D granular unit cell developed in COMSOL MULTIPHYSICS®.}
     \label{DispRel}
\end{figure*}

Figure~\ref{DispRel}(a) illustrates the seismogram obtained from the data collected along the surface green line of Fig.~\ref{ExpSetup}. Several propagation modes arise due to constructive interference phenomena, but mainly two wavetrains dominate the seismogram: a quasi-compressional P wave and a packet of sagittal P-SV waves that propagate with lower velocities. The displacement field associated with these vertically polarized acoustic modes can be described via the following depth-dependent Helmholtz equations:
\begin{equation}
({v_S}^2(z)u'_x)'+\Big((\omega/k)^2-{v_P}^2(z)\Big)u_x = 
i({v_P}^2(z)-2{v_S}^2(z)u'_z+({v_S}^2(z)u_z)')
\label{eq:Helmholtz1}
\end{equation}
\begin{equation}
({v_P}^2(z)u'_z)'+\Big((\omega/k)^2-{v_S}^2(z)\Big)u_z =  i\Big({v_S}^2(z)u'_x+\big(({v_P}^2(z)-2{v_S}^2(z))u_x\big)'\Big)
\label{eq:Helmholtz2}
\end{equation}
where $u_x$ and $u_z$ denote the horizontal and the vertical displacement component, respectively, $\omega$ represents the angular frequency, $k$ the wavenumber, $i$ is the imaginary unit, and the prime encodes the derivative with respect to the vertical ($z$) axis.

\noindent
Figure~\ref{DispRel}(b) shows the frequency-wavenumber spectrum derived after applying the two dimensional (2D) discrete Fourier transform (DFT) to the zero-padded time waveforms. The latter exhibits the frequency content of the acoustic waves detected in the seismogram. The low frequency (0-600 Hz) components of the signal are associated with the two lowest-order P-SV waves (labeled as “P-SV$_1$” and “P-SV$_2$”). It is worth noting that the P-SV$_2$ mode, associated with long-wavelength surface waves, propagates faster than the P-SV$_1$ mode, characterized by shorter wavelengths. Since most of the strain energy associated with surface wave motion is confined within a depth of about one wavelength $\lambda$ from the free boundary \cite{Achenbach}, the P-SV$_2$ mode penetrates deeper into the interior of a medium and, further influenced by the underlying stiffer layers, propagates faster than the P-SV$_1$ mode.\\
The high frequency (600-800 Hz) components of the signal are, instead, associated with the purely compressive (P) vertical wave traveling with higher velocities. These quasi-compressional fast modes can be easily detected in experiments, when the excitation has a significant vertical component of surface displacement \cite{Jacob,Bodet}, as it was the case in our experiment. The displacement field associated with these P modes can be described via Eqs. (\ref{eq:Helmholtz1}-\ref{eq:Helmholtz2}) by neglecting shear rigidity ($v_s$=0) \cite{Jacob}.

\subsection{Numerical dispersion analysis via Finite Element approach}
We confine our analysis to the P-SV acoustic waves and we numerically calculate their dispersion curves by modelling a portion of the granular medium with a Bloch wave - Finite Element (FE) approach, using the software COMSOL Multiphysics®. It has been shown that this numerical procedure allows for properly computing the GSAM dispersion relations \cite{Palermo,Zaccherini2}. We consider a 3D FE model of the unit cell, a column of granular material 1000 mm-high and 30 $\times$ 30 mm wide, as depicted in Fig.~\ref{DispRel}(b). Since the medium is homogeneous along the wave-propagation direction (i.e., along the $x$-axis), the width of the granular column ($D=30$ mm) can be arbitrarily chosen to derive the dispersive properties in the wavenumber range of interest (i.e., from 0 to 100 rad/m, as shown in the experimental $f$-$k$ spectrum illustrated in Fig.~\ref{DispRel}(b)). Since the maximum wavenumber within the range of interest is $k_{max}=100$ rad/m, it follows that $D = \pi/k_{max} \cong 30$ mm.\\
According to the GSAM theory described in Ref. \cite{Aleshin}, the medium is modeled as a linear elastic continuum, assuming constant density and depth-dependent longitudinal and shear velocities profiles: $\displaystyle v_P,_S=\gamma_P,_S(\rho gz)^{\alpha_P,_S}$  with $\gamma_P=14.4$, $\gamma_S=6.42$, and $\alpha_P,_S = 0.31$. We adopted the same elastic power-law coefficients ($\gamma_P,_S$, $\alpha_P,_S$) assumed in Ref. \cite{Palermo} to model the same material. Wave reflections from the bottom of the model are prevented via insertion of absorbing conditions and periodic Bloch boundary conditions are applied to the sides. We perform eigenfrequency analyses by sweeping the wave vector along the Irreducible Brillouin Contour, i.e., by considering the wavenumber spanning in the range of $0 - k_{max}$. The numerical dispersion curves, depicted as solid lines in Fig.~\ref{DispRel}(a), well match the experimental energy distribution of the two P-SV acoustic surface modes.

\section{Wave attenuation mechanisms}
\subsection{Amplitude attenuation of P-SV modes in granular media}
The amplitude of the particle motion associated with the vertically polarized surface waves, generated by a time-harmonic point load, which is applied at the free surface of an unconsolidated granular medium, attenuates not only due to the energy dissipated within the skeletal frame of the medium (i.e., material damping), but also due to the geometric spreading. Consequently, in the far-field, the vertical displacement amplitude induced at the surface by a harmonic point vertical source $F_z \cdot e^{i \omega t}$ in granular media can be expressed as follows \cite{Lai,Rix,Foti}:
\begin{equation}
 \left|U_z(r,{\omega})\right|= F_z\cdot \Upsilon_{P-SV}(r,{\omega},z=0)\cdot e^{-\alpha(\omega) r}
 \label{eq:fitting}
\end{equation}
where $\left|U_z(r,{\omega})\right|$ represents the spectral particle displacement amplitude at the surface, $F_z$ is the amplitude of the excitation, $\Upsilon_{P-SV}(r,{\omega},z=0)$ denotes the function describing the surface geometric spreading of the multimode P-SV waves, and $\alpha(\omega)$, the so-called attenuation coefficient or attenuation constant, is the parameter accounting for the material damping.\\
The geometric spreading function defines how the wave energy spreads into the inhomogeneous medium; in particular, how the P-SV displacement amplitude, for a given frequency, modifies as the wavefront travels away from the source, due to constructive wave interference phenomena occurring between the interfaces of in-depth layers with increasing rigidity. The geometric spreading is mainly controlled by the source type (e.g., point or line source) and the rigidity profile of the medium \cite{Foti}.\\
Instead, the frequency-dependent attenuation constant $\alpha(\omega)$ defines the exponential decay of the wave amplitude caused by the energy dissipated within the soil skeletal frame \cite{Rix}. In geophysics, $\alpha(\omega)$ is a widely used parameter to describe material damping and corresponds to the imaginary part of the complex wavenumber $k*=Re(k*)+Im(k*)=k-i\alpha(\omega)$. Theoretically, at each frequency, there exist as many attenuation coefficients as the total number of modes $M$. However, Eq. \ref{eq:fitting} considers a unique apparent attenuation coefficient that may represent the combination of the several propagating modes \cite{Foti}.\\
The theoretical model represented by Eq. \ref{eq:fitting} independently accounts for the geometric and material attenuation contributions. Since the geometric attenuation depends on the phase velocity, which is, usually, not independent of the material attenuation as a result of material dispersion, we recognize that the geometric attenuation may not be independent from the material one. However, since the numerical dispersion curves, calculated in absence of material damping-induced dissipation, well match the experimental ones, we consider that neglecting this dependence does not significantly affect our analysis.

\subsection{Geometric attenuation law in granular media}
The geometric spreading function of the GSAMs is computed using the explicit solution of the geometric attenuation law of Rayleigh-like waves in layered media, i.e., vertically heterogeneous media composed of multiple homogeneous linear elastic layers overlaying a homogeneous half-space. Indeed, the P-SV surface acoustic modes display a polarization in the vertical plane similar to that of the Rayleigh wave and the granular substrate featuring the power-law velocity profile can be considered as a stratified medium composed of infinitesimally thin layers. In addition, the dynamics of the first sagittal mode $P-SV_1$, which carries most of the energy in our experiment, resembles that of the Rayleigh wave in homogeneous media, tending to it in the presence of strong particle adhesion \cite{Aleshin}.

\subsubsection{Geometric spreading functions computed via modal superposition}
\begin{figure*}
\includegraphics[width=1\textwidth]{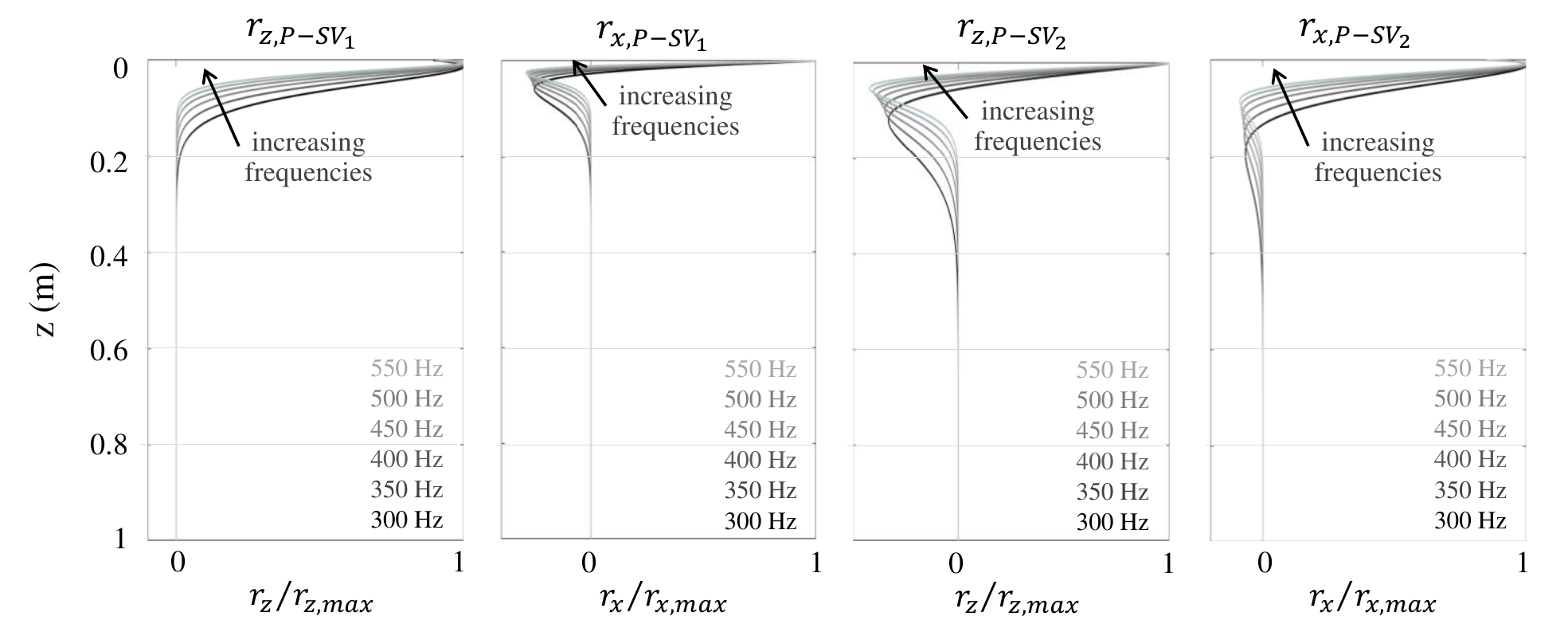}
  \caption{Numerical mode shapes. Vertical ($r_{z,P-SV_1}$, $r_{z,P-SV_2}$) and horizontal ($r_{x,P-SV_1}$, $r_{x,P-SV_2}$) displacement component of the first and second lowest-order P-SV acoustic surface modes, respectively.}
     \label{ModeShape}
\end{figure*}

In lossless, vertically layered media, the far field wave motion induced by a time-harmonic, vertical point load applied at the free surface results via the superposition of several propagation modes each traveling, for a given frequency, at a different phase velocity. These multiple modes originate from interference phenomena arising among waves undergoing multiple reflections and refractions at the layer interfaces. The geometric attenuation law $\Upsilon(r,{\omega},z)$ resulting from the superposition of $M$ different Rayleigh-like modes can be expressed as follows \cite{Lai}:
\begin{equation}
\Upsilon(r,{\omega},z)=\frac{1}{4\sqrt{2\pi r}} \bigg\{\sum_{i=1}^M \sum_{j=1}^M \frac{r_z(k_i,\omega,z)r_z(k_j,\omega,z)r_z(k_i,\omega,z=0)r_z(k_j,\omega,z=0)\cos[r(k_i-k_j)]}{\sqrt{k_ik_j}[V_iU_iI_i][V_jU_jI_j]} \bigg\}^{0.5}
\label{eq:Spreading}
\end{equation}
where $r_z(k_j,\omega,z)$ and $r_x(k_j,\omega,z)$ are the vertical and the horizontal eigenfunctions associated with the $j$-th mode of propagation characterized by a wavenumber $k_j$. $V_j$ and $U_j$ are the phase and group velocity evaluated at $\omega$-$k_j$, whereas $I_j$ represents the Rayleigh-like energy integral \cite{Aki} evaluated at $\omega$-$k_j$ and defined as:
\begin{equation}
 I_j(k_j,\omega,z)=\frac{1}{2} \int_{0}^{\infty}\rho(z)[{r_x^2(k_j,\omega,z)}+{r_z^2(k_j,\omega,z)}] dz
 \label{eq:energy}
\end{equation}
\noindent
The total number of modes $M$ required to properly derive the surface geometric spreading function depends on the nature of the elastic profile of the soil, i.e., how the shear modulus varies with depth, and on the excitation frequency. Here, the spatial attenuation law is calculated considering the first two lowest-order P-SV modes, which dominate the experimental frequency-wavenumber spectrum illustrated in Fig.~\ref{DispRel}(a), hence, $M=2$. Since the main frequency content of these modes extends from 300 to 550 Hz, we compute the geometric spreading function at the surface ($z=0$), for values of $r$ ranging from 0 to 1 m, and for six equispaced frequency values within the aforementioned frequency range (i.e., at 300, 350, 400, 450, 500, 550 Hz). All input parameters including wavenumbers, eigenfunctions, phase and group velocities, and Rayleigh-like energy integral are obtained numerically from the eigenfrequency study performed on the granular unit cell. Figure~\ref{ModeShape} illustrates the vertical ($z$) and horizontal ($x$) displacement profiles of the two P-SV modes associated with the six aforementioned frequency values. 

\begin{figure*}
\includegraphics[width=1\textwidth]{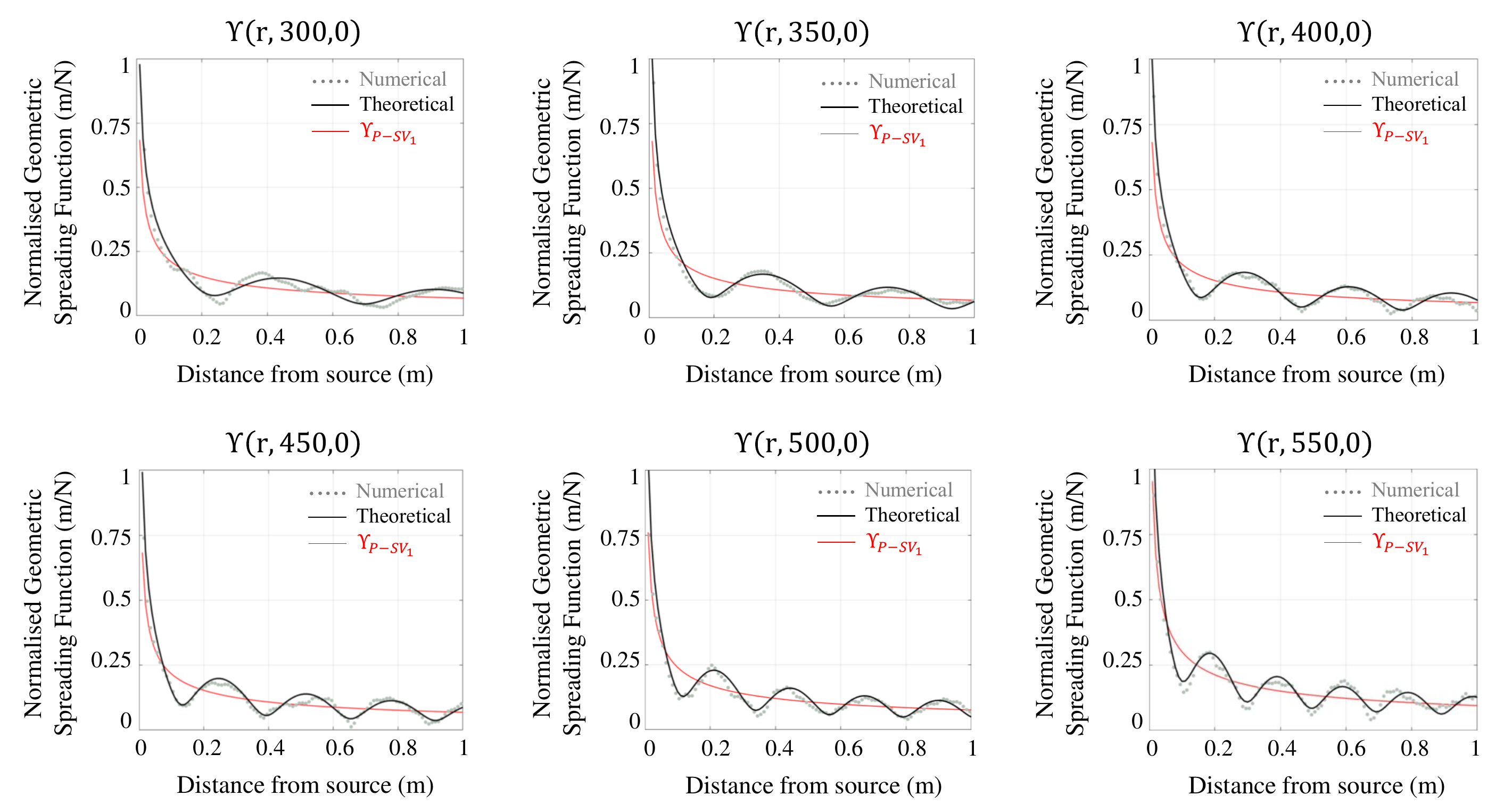}
  \caption{Geometric spreading functions computed at the surface ($z=0$) at 300, 350, 400, 450, 500, and 550 Hz, using the explicit solution of the geometric attenuation law of Rayleigh-like waves in stratified media (solid black lines) and performing numerical time-domain simulations (grey dots). The solid red lines represent the geometric spreading functions $\Upsilon_{P-SV_1}$ calculated considering only the contribution of the fundamental P-SV$_1$ mode.}
     \label{GeomSpread}
\end{figure*}

\noindent
The maximum displacement occurs in the near-surface layers, consistently for any surface mode and the wave energy, localized at the surface, decays rapidly with depth. As expected, the mode penetration depth, which is typically roughly equal to one wavelength $\lambda$, increases for decreasing frequency values. Furthermore, as disclosed in Refs. \cite{Gusev,Aleshin}, the number of “phases” in the displacement profile, as a function of the depth, corresponds to the order of modes.\\
The computed geometric spreading functions $\Upsilon(r,{\omega},z=0)$ describing the spatial attenuation of the P-SV modes at the surface are depicted as solid black lines in Fig.~\ref{GeomSpread}.

\noindent
The interference among the acoustic modes produces pronounced oscillations in the spreading functions that, as revealed in Ref. \cite{Foti}, multiply for increasing frequencies. These oscillations are not present in the simple classical attenuation law of the Rayleigh wave in homogeneous media proportional to $1/\sqrt{r}$ or in the case of single-mode wave propagation (see the solid red lines in Fig.~\ref{GeomSpread} depicting the spreading functions, labelled here $\Upsilon_{P-SV_1}$, calculated considering only the contribution of the fundamental P-SV$_1$ mode). Indeed, if we take into account only the contribution of the P-SV$_1$ mode, then $M=1$, $k_i=k_j=k_1$, $V_i=V_j=V_1$, $U_i=U_j=U_1$, $I_i=I_j=I_1$, and the geometric spreading function reduces to the relation $\Upsilon_{P-SV_1}= E_z / \sqrt{r} $, with $E_z= r_z(z,\omega) r_z(0,\omega) / 4 V_1 U_1 I_1 \sqrt{2\pi k}$.

\subsubsection{Geometric spreading functions computed by means of 3D numerical simulations}
\begin{figure*}
\includegraphics[width=1\textwidth]{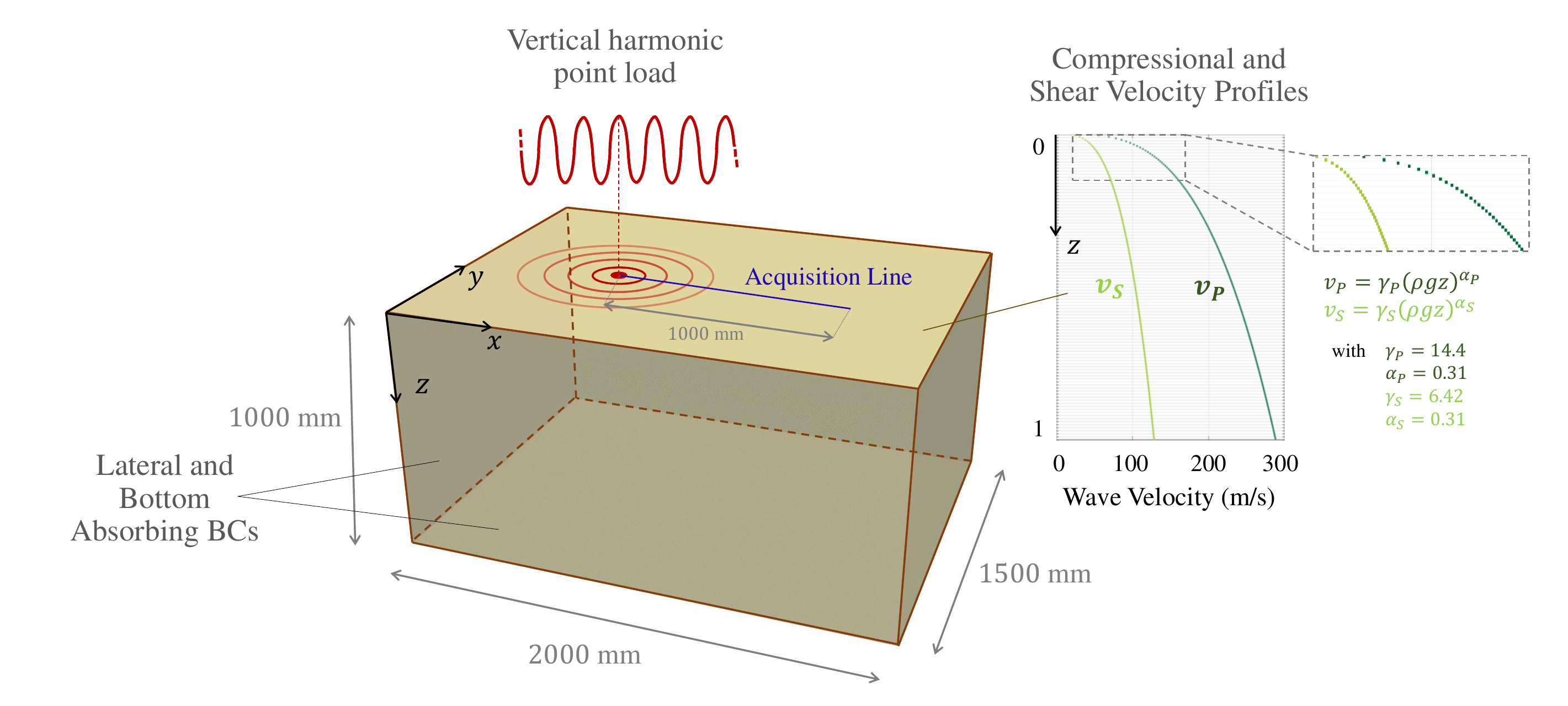}
  \caption{Numerical time domain simulation. The computational domain characterized by the granular substrate excited by a vertical, time-harmonic point load applied at the free surface.}
     \label{model}
\end{figure*}

Next, we compare the surface geometric spreading functions calculated via modal superposition with the wave displacement amplitude decay computed by means of three-dimensional (3D) time-transient numerical simulations performed using SPECFEM3D \cite{Komatitsch}, a spectral element-based software able to solve large-scale time-dependent elastodynamic problems. In particular, we consider a 1000-mm-deep granular half-space, 2000 $\times$ 1500 mm wide, modeled as a linear elastic continuum, defined by the previously assumed power-law rigidity profile, according to the GSAM theory described in Ref. \cite{Aleshin} (see Fig.~\ref{model}).

\noindent
A mesh of hexahedral elements is generated using the software TRELIS 16.5 through PYTHON scripting. Since the velocity profiles are discretized accordingly to the discretization of the computational domain, we choose the element size equal to 5 mm to accurately model the exponential increase of $v_s$ and $v_s$, especially in the near-surface layers. The inset on the right depicts the adopted discretization of the compressional and shear velocity profiles. Furthermore, the GSAM theory establishes that, in absence of adhesion, the compressional and shear velocities vanish at the free surface. However, the employed computational software (SPECFEM3D) requires non-vanishing velocity values in each of the mesh points. Therefore, we impose at the free surface velocity values calculated for a depth $z$ equal to 2 mm.\\
A vertical, time-harmonic point load applied at the free surface and located at a distance of 300 mm from the left edge, generates vertically polarized modes. To avoid reflections from the boundaries, we apply absorbing conditions to the bottom and side edges.\\
The spatial attenuation laws of the P-SV modes obtained by means of 3D numerical simulations are depicted as dotted grey lines in Fig.~\ref{GeomSpread}. These are derived by computing the vertical component of the wave displacement at the surface, along the 1-m long line illustrated in blue in Fig.~\ref{model}. The wave displacement amplitude decays computed by means of 3D numerical simulations (dotted grey lines) well match the geometric spreading functions calculated via modal superposition (solid black lines). This confirms that the analytical technique is able to correctly describe the P-SV wave displacement amplitude as well as capture its oscillations at different frequencies.

\subsection{Material damping-induced dissipation}
\begin{figure*}
\includegraphics[width=1\textwidth]{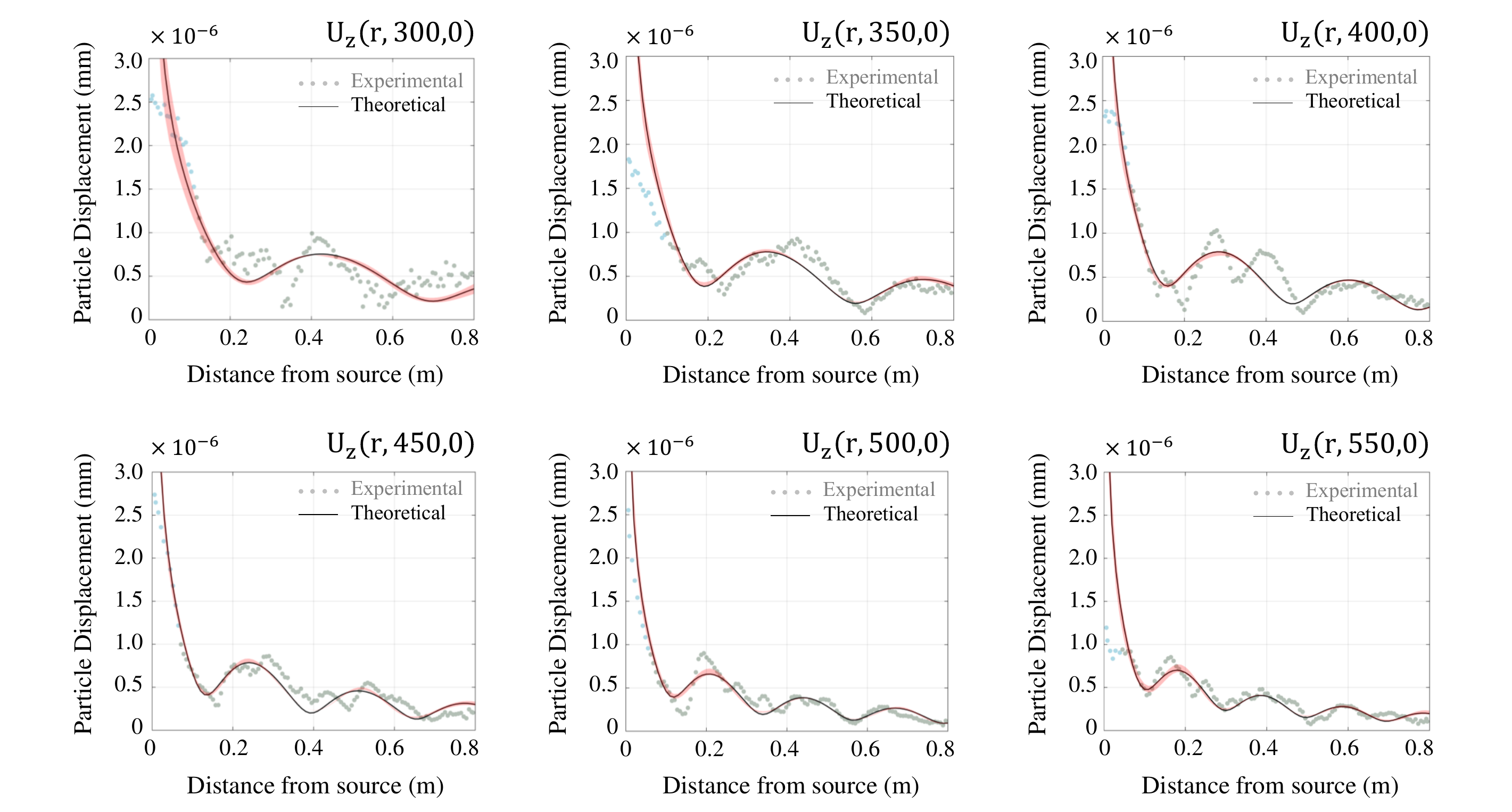}
  \caption{Spectral P-SV surface wave displacement. Experimental P-SV wave spectral displacement (grey dots) and theoretical displacement functions (solid black lines) resulting from the non-linear regression analysis at 300, 350, 400, 450, 500, and 550 Hz. With yellow dots we depict the displacement data discarded in the regression analysis because located in the near field. The upper and lower bounds of the confidence intervals define the boundaries of 95$\%$ confidence region, which is marked as a red shaded area.}
     \label{expdata}
\end{figure*}

To determine the frequency-dependent attenuation coefficients, we consider the data collected in the laboratory experiment. In particular, we compute the vertical displacement time histories from the velocity data recorded at the surface along the green line illustrated in Fig.~\ref{ExpSetup}. The numerical integration is performed using a simple trapezoidal rule, after employing a linear high-pass filter in order to avoid signal drifts. For each acquisition point along the surface line, we calculate the surface spectral displacement amplitude $\left|U_z(r,{\omega})\right|$ associated with each of the six frequency values previously considered\footnote{In the experiment, the vertical motion is realized by means of a rod exciting the granular medium with an angle of 20$^\circ$ from the normal to the free surface. As a result, the excitation features a small ($F_x=F_z \cdot \tan (20^\circ)$) horizontal component along the $x$-axis. Therefore, we verify that the presence of $F_x$ does not significantly affect our analysis. In particular, we perform 3D numerical simulations with and without the presence of $F_x$ using SPECFEM3D and we numerically compute the spectral coherence between the time waveforms obtained from the two simulations.}. The results are depicted as grey dots in Fig.~\ref{expdata}.

\begin{figure}
\includegraphics[width=0.5\textwidth]{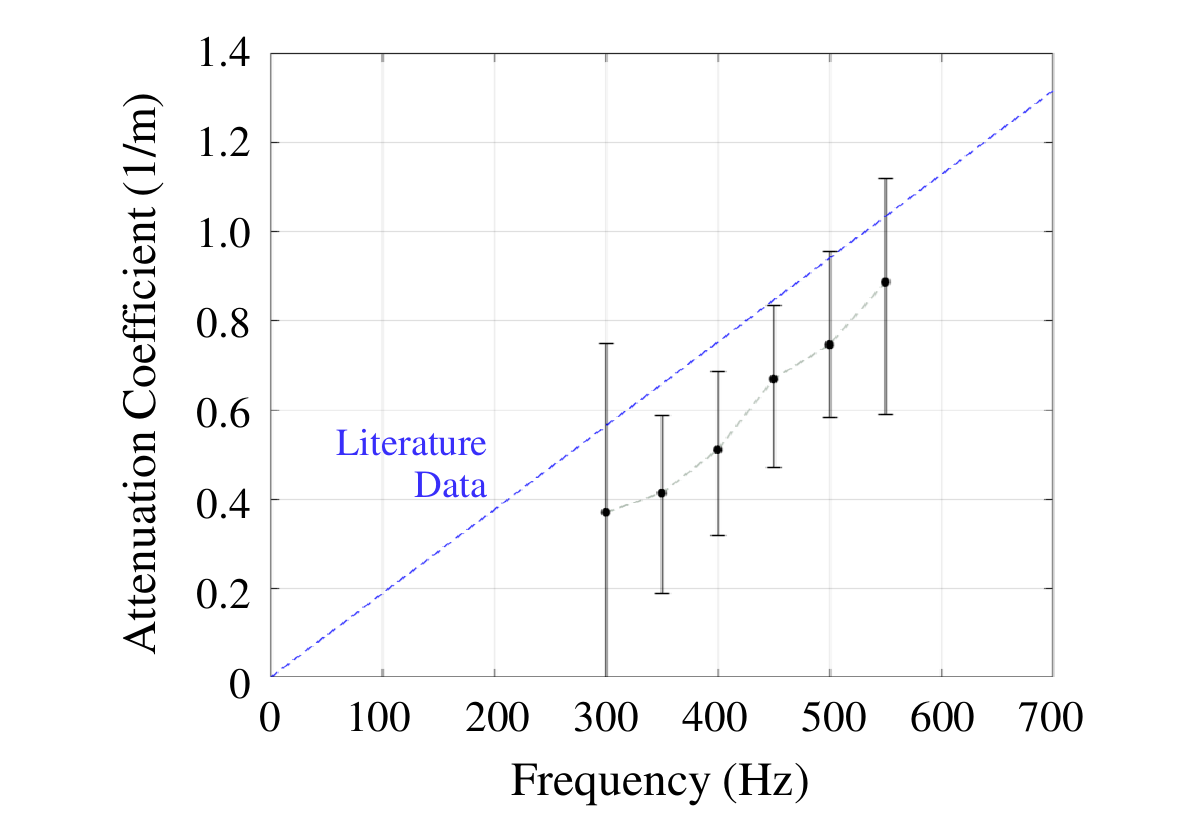}
  \caption{P-SV wave attenuation curve. Attenuation curve of P-SV surface acoustic waves propagating at the surface of a granular medium composed of 150 ${\mu} $m-diameter silica microbeads. The obtained attenuation values are depicted as black dots together with their confidence intervals, while the blue line represents the trend of shear attenuation data found in literature for 105-210 ${\mu} $m-diameter glass beads.}
     \label{attcurve}
\end{figure}

\noindent
Then, according to Eq. \ref{eq:fitting}, we perform a non-linear regression analysis on the experimental data $\left|U_z(r,{\omega})\right|$ to derive the material attenuation coefficients $\alpha_{mat}(\omega)$. We discard from the analysis the acquisition points lying at a distance less than $\lambda_{P-SV}/2$ from the source, with $\lambda_{P-SV}$ representing the wavelength of the P-SV waves (see the yellow points in Fig.~\ref{expdata}). Indeed, numerical studies of surface wave propagation in vertically inhomogeneous soil profiles \cite{Holzlohner,Vrettos,Tokimatsu} have shown that, in normally dispersive media, i.e., media with a mechanical impedance $\rho v_s$ that is constant or smoothly increases with depth, the near-field (body wave) effects become negligible and, consequently, the wavefield is dominated by surface waves when the distance from the source exceeds about $\lambda/2$.
Aside from the material attenuation constant $\alpha_{mat}(\omega)$, we also consider the excitation magnitude $F_z$ as an unknown parameter, in order to alleviate uncertainties due to coupling between the source and the medium \cite{Foti}. The regression analysis yields a series of material attenuation coefficients, shown in Fig.~\ref{attcurve} and collected in Table~\ref{symbols}.

\begin{table}
\caption{Frequency-dependent material attenuation coefficients obtained from the regression analysis that was carried out on the experimental displacement data. The latter are computed from the velocity data that has been recorded along the surface green line, illustrated in Fig.~\ref{ExpSetup}.}
\label{symbols}
\begin{tabular}{@{}lllllll}
$f$ (Hz) & 300 & 350 & 400 & 450 & 500 & 550\\
\hline
$\alpha_{mat}$ (1/m) & 0.370 & 0.413 & 0.510 & 0.669 & 0.746 & 0.887\\
\hline
\end{tabular}
\end{table}

\noindent
As expected, the material attenuation coefficient $\alpha_{mat}(\omega)$ increases with frequency. The theoretical fitted functions, represented as solid black lines in Fig.~\ref{expdata}, well match the experimental displacement data, thoroughly capturing the oscillations arising as a result of the mode superposition. However, there exist additional oscillations that are not reproduced by the theoretical model, but are visible in the experimental data, in particular at 300 and 400 Hz. These oscillations may be attributed to the influence of the pure compressional wave. Indeed, by observing the experimental dispersion curve (Fig.~\ref{DispRel}(a)), we note that part of the energy of the P wave is localized at low frequencies, particularly for the values of 300 and 400 Hz, thus justifying this mismatch.\\
The 95$\%$ confidence intervals for the coefficients $\alpha_{mat}$ of the nonlinear regression model, computed using the bootstrap method, are depicted in black in Fig.~\ref{attcurve}. The upper and lower bounds of the confidence intervals define the boundaries of 95$\%$ confidence region, which is marked as a red shaded area in Fig.~\ref{expdata}.\\
To validate the findings, our results are compared to attenuation values found in literature for unconsolidated granular materials. Experimental studies \cite{Bell} report shear wave attenuation measurements in 105-210 ${\mu} $m-diameter dry glass beads for frequencies ranging from 500 Hz to 20 kHz. Although the P-SV modes do not propagate within the material bulk as shear waves, but in the near-surface layers, similar attenuation coefficients may be expected owing to their predominant shear component. The reported data, which was subsequently validated via the grain-shearing theory, described in Ref. \cite{Buckingham}, indicates a linear frequency dependence (represented by the blue dashed line in Fig.~\ref{attcurve}), which proves consistent with the attenuation coefficients $\alpha_{mat}$ obtained from the regression analysis. In addition, the linear frequency dependence of the attenuation coefficients, predicted by the grain-shearing theory, indicates that the results of this work can be extended to lower frequencies.\\
Finally, note that laboratory and field tests on cohesionless sandy soils \cite{Seed} suggest values of damping ratios consistent with those obtained for our granular medium.

\section{Conclusions}
In conclusion, we experimentally and numerically analyse the dynamics and the attenuation of vertically polarized acoustic waves propagating at the surface of an unconsolidated granular medium, which features a power-law stiffness profile. Dispersion analysis reveals that the surface wavefield is dominated by the two lowest-order P-SV modes, whose frequency content mainly extends from 300 to 550 Hz. The existence, for a given frequency, of multiple vertically polarized acoustic modes, each characterized by its own vertical and horizontal displacement profile along the depth, is due to constructive interference phenomena occurring in the inhomogeneous granular material.\\
Subsequently, we analyse the particle displacement amplitude decay, investigating the two main attenuation mechanisms that contribute to wave energy dissipation during dynamic excitation, namely the geometric spreading and the material damping. First, we calculate the GSAM geometric spreading function by adopting an approach originally developed to quantify the spatial attenuation law of seismic Rayleigh-like waves in inhomogeneous stratified media. In virtue of the linearity of these surface acoustic modes, the medium is modeled as perfectly continuous and elastic, despite its granular nature. As an effect of the mode superposition, the obtained geometric spreading functions exhibit pronounced oscillations that multiply for increasing frequencies. Then, frequency-dependent attenuation coefficients are determined by means of non-linear regression analysis, which was carried out on experimental particle displacement data, to assess the intrinsic energy dissipation induced by material damping. The outcomes of the nonlinear regression reveal attenuation coefficient values increasing with the frequency. The latter proves consistent with attenuation values found in literature for dry glass beads with similar grain size and damping ratios of cohesionless sandy soils.\\
In addition, the pronounced amplitude oscillations of the experimental particle displacement are well captured by the theoretical spatial attenuation law of seismic Rayleigh-like waves in inhomogeneous stratified media. Firstly, this demonstrates that the continuous medium approximation can be applied not only to estimate the elastic properties of the material or investigate the GSAM dispersive properties (as the previous studies show), but further to determine the wave geometric spreading caused by the medium inhomogeneity and the attenuation coefficients accounting for the energy dissipated between the grains. Furthermore, it confirms that techniques that are commonly used at the geophysics scale prove successful also when applied to laboratory-scaled physical models.\\
Future studies should investigate how the attenuation coefficient value is affected upon increase of the adhesion between grains. This can be achieved by gradually injecting various types of pore fluid into the granular layer in a controlled manner \cite{Bodet}. In the presence of strong adhesion, the medium is expected to become almost homogeneous near the surface. The lowest-order mode should transform into the Rayleigh mode, while the further surface modes delocalize. As a result, the GSAM geometric spreading should gradually tend to the classical attenuation law of the Rayleigh wave (the amplitude oscillations should disappear) and the attenuation coefficient value increase.

\begin{acknowledgments}
This work was partially supported by the ETH Research Grant (49 17-1) to E.C.
\end{acknowledgments}

\begin{dataavailability}
The data that support the findings of this study are available
from the corresponding author upon reasonable request.
\end{dataavailability}

\label{lastpage}


\begin{thebibliography}{}
    
  \bibitem[\protect\citename{Achenbach }1984]{Achenbach}
    Achenbach, J. D., 1984. \textit{Wave Propagation in Elastic Solids}, Amsterdam, The Netherlands: North-Holland Publishing.
    
  \bibitem[\protect\citename{Aki }1980]{Aki}
    Aki, K. \& Richards, P. G., 1980. \textit{Quantitative seismology: Theory and methods}, W. H. Freeman and Co., San Francisco.
    
  \bibitem[\protect\citename{Aleshin }2007]{Aleshin}
      Aleshin, V., Gusev, V., \& Tournat, V., 2007. Acoustic modes propagating along the free surface of granular media,
    \textit{J. Acoust. Soc. Am.}, \textbf{121}, 2600.
    
  \bibitem[\protect\citename{Askan }2007]{Askan}
      Askan, A., Akcelik, V., Bielak, J., \& Ghattas, O., 2007. Full Waveform Inversion for Seismic Velocity and Anelastic Losses in Heterogeneous Structures,
    \textit{Bull. Seismological Soc. of Am.}, \textbf{97}, 1990-2008.
    
  \bibitem[\protect\citename{Badsar }2010]{Badsar}
     Badsar, S. A., Schevenels, M., Haegeman, W., \& Degrande, G., 2010. Determination of the material damping ratio in the soil from SASW tests using the half-power bandwidth method,
    \textit{Geophys. J. Int.}, \textbf{182}, 1493-1508.
    
   \bibitem[\protect\citename{Bell }1979]{Bell}
    Bell, D. W., 1979, \textit{Technical Report No. ARL-TR-79-31}, Applied Research Laboratories, The University of Texas at Austin.
    
  \bibitem[\protect\citename{Bergamo }2014]{Bergamo}
      Bergamo, P., Bodet, L., Socco, L. V., Mourgues, R., \& Tournat V., 2014. Physical modelling of a surface-wave survey over a laterally varying granular medium with property contrasts and velocity gradients,
    \textit{Geophys. J. Int.}, \textbf{197}, 233-247.
    
  \bibitem[\protect\citename{Bodet }2009]{Bodet4}
      Bodet, L., Abraham, O. \& Clorennec, D., 2009. Near-offset effects on Rayleigh-wave dispersion measurements: Physical modeling,
    \textit{J. Appl. Geophy.}, \textbf{68}, 95-103.
    
  \bibitem[\protect\citename{Bodet }2014]{Bodet}
      Bodet, L., Dhemaied, A., Martin, R., Mourgues, R., Rejiba, F., \& Tournat, V., 2014. Small-scale physical modeling of seismic-wave propagation using unconsolidated granular media,
    \textit{Geophysics}, \textbf{79}, T323.
    
  \bibitem[\protect\citename{Bodet }2010]{Bodet2}
      Bodet, L., Jacob, X., Tournat, V., Mourgues, R. \& Gusev, V., 2010. Elasticity profile of an unconsolidated granular medium inferred from guided waves: Toward acoustic monitoring of analogue models,
    \textit{Tectonophysics}, \textbf{496}, 99-104.
    
  \bibitem[\protect\citename{Bodet }2005]{Bodet3}
      Bodet, L., van Wijk, K., Bitri, A., Abraham, O., C\^{o}te, P., Grandjean, G. \& Leparoux, D., 2005. Surface-wave inversion limitations from laser-Doppler physical modeling,
    \textit{J. Environ. Eng. Geophys.}, \textbf{10}, 151-162.
    
  \bibitem[\protect\citename{Bonneau }2007]{Bonneau}
      Bonneau, L., Andreotti, B., \& Clément, E., 2007. Surface elastic waves in granular media under gravity and their relation to booming avalanches,
    \textit{Phys. Rev. E}, \textbf{75}, 016602.
    
  \bibitem[\protect\citename{Bretaudeau }2011]{Bretaudeau}
      Bretaudeau, F., Leparoux, D., Durand, O. \& Abraham, O., 2011. Small-scale modeling of onshore seismic experiment: A tool to validate numerical modeling and seismic imaging methods,
    \textit{Geophysics}, \textbf{76}, T101-T112.
    
   \bibitem[\protect\citename{Buckingham }2014]{Buckingham}
    Buckingham, M. J., 2014. Analysis of shear-wave attenuation in unconsolidated sands and glass beads,
    \textit{J. Acoust. Soc. Am.}, \textbf{136}, 2478.
    
   \bibitem[\protect\citename{Campman }2005]{Campman}
    Campman, X. H., Van Wijk, K., Scales, J. A. \& Herman, G. C., 2005. Imaging and suppressing near-receiver scattered surface waves,
    \textit{Geophysics}, \textbf{70}, V21-V29.
    
   \bibitem[\protect\citename{Dewangan }2006]{Dewangan}
    Dewangan, P., Tsvankin, I., Batzle, M., Van Wijk, K. \& Haney, M., 2006. PS-wave moveout inversion for tilted TI media: A physical-modeling study,
    \textit{Geophysics}, \textbf{71}, D135-D143.
    
    \bibitem[\protect\citename{Field }1993]{Field}
    Field, E. H., \& Klaus H. J., 1993. "Monte-Carlo simulation of the theoretical site response variability at Turkey Flat, California, given the uncertainty in the geotechnically derived input parameters,
    \textit{Earthquake Spectra}, \textbf{9}, 669-701.
    
  \bibitem[\protect\citename{Foti }2015]{Foti}
    Foti, S., Lai, C. G., Rix, G. J., \& Strobbia, C., 2015. \textit{Surface Wave Methods for Near-Surface Site Characterization}, International Standard Book.
    
  \bibitem[\protect\citename{Gassmann }1951]{Gassmann}
      Gassmann, F., 1951. Elastic waves through a packing of spheres,
    \textit{Geophysics}, \textbf{16}, 673-685.

  \bibitem[\protect\citename{Gibbs }1994]{Gibbs}
      Gibbs, J. F., Boore, D. M., Joyner, W. B., \& Fumal, T. E., 1994. The attenuation of seismic shear waves in quaternary alluvium in Santa Clara Valley, California,
    \textit{Bull. Seismological Soc. of Am.}, \textbf{84}, 76-90.
    
  \bibitem[\protect\citename{Gusev }2006]{Gusev}
      Gusev, V. E., 2006. Acoustic Waves in an Elastic Channel near the Free Surface of Granular Media,
    \textit{Phys. Rev. Lett.}, \textbf{96}, 214301.
    
  \bibitem[\protect\citename{Gusev }2008]{Gusev2}
      Gusev, V. E., \& Tournat, V., 2008. How acoustic waves are guided in buried subsurface channels in unconsolidated granular media,
    \textit{Phys. Rev. E}, \textbf{78}, 036602.
    
   \bibitem[\protect\citename{Holzlohner }1980]{Holzlohner}
    Holzlohner, U., 1980. Vibrations of the elastic half-space to vertical surface loads,
    \textit{Earthquake Engrg. and Struct. Dyn.}, \textbf{8}, 405-414.
    
  \bibitem[\protect\citename{Jacob }2008]{Jacob}
      Jacob, X., Aleshin, V., Tournat, V., Leclaire, P., Lauriks, W., \& Gusev, V. E., 2008. Acoustic Probing of the Jamming Transition in an Unconsolidated Granular Medium,
    \textit{Phys. Rev. Lett.}, \textbf{100}, 158003.
    
  \bibitem[\protect\citename{Jongmans }1990]{Jongmans}
      Jongmans, D., 1990. In-situ attenuation measurements in soils,
    \textit{Engrg. Geology}, \textbf{29}, 99-118.
    
    \bibitem[\protect\citename{Komatitsch }1998]{Komatitsch}
    Komatitsch, D. \& Vilotte, J.-P., 1998. The spectral element method: An efficient tool to simulate the seismic response of 2D and 3D geological structures,
    \textit{Bull. Seismological Soc. of Am.}, \textbf{88}, 368-392.
    
    \bibitem[\protect\citename{Krawczyk }2013]{Krawczyk}
    Krawczyk, C. M., Buddensiek, M. L., Oncken, O. \& Kukowski, N., 2013. Seismic imaging of sandbox experiments–laboratory hardware setup and first reflection seismic sections,
    \textit{Solid Earth}, \textbf{4}, 93-104.
    
  \bibitem[\protect\citename{Lai }1998]{Lai}
    Lai, C. G. \& Rix, G. J., 1998.  Simultaneous inversion of Rayleigh phase velocity and attenuation for near-surface site characterization, \textit{PhD thesis}, 
    N.S.F. and U.S.G.S., Georgia Institute of Technology.
    
  \bibitem[\protect\citename{Liu }1992]{Liu}
      Liu, C-H., \& Nagel, S. R., 1992. Sound in sand,
    \textit{Phys. Rev. Lett.}, \textbf{68}, 2301–2304.
    
  \bibitem[\protect\citename{Makse }2004]{Makse}
      Makse, H. A., Gland, N., Johnson, D. L., \& Schwartz, L., 2004. Granular packings: Nonlinear elasticity, sound propagation, and collective relaxation dynamics,
    \textit{Phys. Rev. E}, \textbf{70}, 061302.
    
  \bibitem[\protect\citename{Palermo }2018]{Palermo}
     Palermo, A., Kr\"odel, S., Matlack, K. H., Zaccherini, R., Dertimanis, V. K., Chatzi, E. N., Marzani, A., \& Daraio, C., 2018. Hybridization of Guided Surface Acoustic Modes in unconsolidated granular media by a resonant metasurface,
    \textit{Phys. Rev. Appl.}, \textbf{9}, 054026.

  \bibitem[\protect\citename{Pasquet }2015]{Pasquet}
      Pasquet, S., Bodet, L., Vitale, Q., Rejiba, F., Guérin, R., Mourgues, R., \& Tournat, V., 2015. Laser-Doppler acoustic probing of granular media with varying water levels,
    \textit{Physics Procedia}, \textbf{70}, 799.

  \bibitem[\protect\citename{Rix }2000]{Rix}
      Rix, G. J., Lai, C. G., \& Spang, A. W., 2000. In Situ Measurement of Damping Ratio Using Surface Waves,
    \textit{J. Geotech. Geoenviron. Eng.}, \textbf{126}, 472-480.
    
  \bibitem[\protect\citename{Seed }1986]{Seed}
      Seed, H. B., Wong, R. T., Idriss, I. M., \& Tokimatsu, K., 1986. Moduli and damping factors for dynamic analyses of cohesionless soils,
    \textit{Int. J. Geotech. Eng.}, \textbf{112}, 1016-1032.
    
  \bibitem[\protect\citename{Sherlock }2001]{Sherlock}
      Sherlock, D. H. \& Evans, B. J., 2001. The development of seismic reflection sandbox modeling,
    \textit{Am. Assoc. Pet. Geol. Bull.}, \textbf{85}, 1645-1659.
    
  \bibitem[\protect\citename{Stewart }1992]{Stewart}
    Stewart, W. P., 1992.  In situ measurement of dynamic soil properties with emphasis on damping, \textit{PhD thesis}, 
    University of British Columbia, Vancouver.
    
    \bibitem[\protect\citename{Tokimatsu }1997]{Tokimatsu}  Tokimatsu, K., 1997. Geotechnical site characterization using surface waves,
   \textit{Earthquake Geotechnical Engineering}, \textbf{3}, 1333-1368
  
  \bibitem[\protect\citename{Verachtert }2018]{Verachtert}
     Verachtert, R., Lombaert, G., \& Degrande, G., 2018. Multimodal determination of Rayleigh dispersion and attenuation curves using the circle fit method,
    \textit{Geophys. J. Int.}, \textbf{212}, 2143-2158.
    
   \bibitem[\protect\citename{Vrettos }1991]{Vrettos}
    Vrettos, C., 1991. Time-harmonic Boussinesq problem for a continuously non-homogeneous soil,
    \textit{Earthquake Engrg. and Struct. Dyn.}, \textbf{20}, 961-977.
    
  \bibitem[\protect\citename{Zaccherini }2020]{Zaccherini1}
     Zaccherini, R., Colombi, A., Palermo, A., Dertimanis, V. K., Marzani, A., Thomsen, H. R., Stojadinovic, B., \& Chatzi, E., 2020. Locally Resonant Metasurfaces for Shear Waves in Granular Media,
    \textit{Phys. Rev. Appl.}, \textbf{13}, 034055.
    
  \bibitem[\protect\citename{Zaccherini }2020]{Zaccherini2}
     Zaccherini, R., Palermo, A., Marzani, A., Colombi, A., Dertimanis, V., \& Chatzi, E., 2020. Mitigation of Rayleigh-like waves in granular media via multi-layer resonant metabarriers,
    \textit{Appl. Phys. Lett.}, \textbf{17}, 254103.

\end{thebibliography}
\end{document}